\documentclass[aps,prl,twocolumn,superscriptaddress,showpacs]{revtex4}
\usepackage{mathrsfs}
\usepackage{amssymb}
\usepackage{txfonts}
\usepackage{tipa}

\usepackage{graphicx}
\usepackage{dcolumn}
\usepackage{bm}

\begin{document}


\title{Magnetic anisotropic energy gap and low energy spin wave excitation

in antiferromagnetic block phase of K$_{2}$Fe$_{4}$Se$_{5}$}

\author{Y. Xiao}
\email[y.xiao@fz-juelich.de]{}
\affiliation{J\"{u}lich Centre for Neutron Science JCNS and Peter Gr\"{u}nberg Institut PGI, JARA-FIT, Forschungszentrum J\"{u}lich GmbH, D-52425 J\"{u}lich, Germany}

\author{S. Nandi}
\affiliation{J\"{u}lich Centre for Neutron Science JCNS and Peter Gr\"{u}nberg Institut PGI, JARA-FIT, Forschungszentrum J\"{u}lich GmbH, D-52425 J\"{u}lich, Germany}

\author{Y. Su}
\affiliation{J\"{u}lich Centre for Neutron Science JCNS-FRM II, Forschungszentrum J\"{u}lich GmbH, Outstation at FRM II, Lichtenbergstra{\ss}e 1, D-85747 Garching, Germany}

\author{S. Price}
\affiliation{J\"{u}lich Centre for Neutron Science JCNS and Peter Gr\"{u}nberg Institut PGI, JARA-FIT, Forschungszentrum J\"{u}lich GmbH, D-52425 J\"{u}lich, Germany}

\author{H.-F. Li}
\affiliation{J\"{u}lich Centre for Neutron Science, Forschungszentrum J\"{u}lich, Outstation at Institut Laue-Langevin, BP 156, 38042 Grenoble Cedex 9, France}

\author{Z. Fu}
\affiliation{J\"{u}lich Centre for Neutron Science JCNS-FRM II, Forschungszentrum J\"{u}lich GmbH, Outstation at FRM II, Lichtenbergstra{\ss}e 1, D-85747 Garching, Germany}

\author{W. Jin}
\affiliation{J\"{u}lich Centre for Neutron Science JCNS and Peter Gr\"{u}nberg Institut PGI, JARA-FIT, Forschungszentrum J\"{u}lich GmbH, D-52425 J\"{u}lich, Germany}

\author{A. Piovano}
\affiliation{Institut Laue-Langevin, 6 rue Jules Horowitz, 38042 Grenoble Cedex 9, France}

\author{A. Ivanov}
\affiliation{Institut Laue-Langevin, 6 rue Jules Horowitz, 38042 Grenoble Cedex 9, France}

\author{K. Schmalzl}
\affiliation{J\"{u}lich Centre for Neutron Science, Forschungszentrum J\"{u}lich, Outstation at Institut Laue-Langevin, BP 156, 38042 Grenoble Cedex 9, France}

\author{W. Schmidt}
\affiliation{J\"{u}lich Centre for Neutron Science, Forschungszentrum J\"{u}lich, Outstation at Institut Laue-Langevin, BP 156, 38042 Grenoble Cedex 9, France}

\author{T. Chatterji}
\affiliation{Institut Laue-Langevin, 6 rue Jules Horowitz, 38042 Grenoble Cedex 9, France}

\author{Th. Wolf}
\affiliation{Institut f\"{u}r Festk\"{o}rperphysik, Karlsruhe Institute of Technology, D-76021 Karlsruhe, Germany}
\author{Th. Br\"{u}ckel}
\affiliation{J\"{u}lich Centre for Neutron Science JCNS and Peter Gr\"{u}nberg Institut PGI, JARA-FIT, Forschungszentrum J\"{u}lich GmbH, D-52425 J\"{u}lich, Germany}
\affiliation{J\"{u}lich Centre for Neutron Science JCNS-FRM II, Forschungszentrum J\"{u}lich GmbH, Outstation at FRM II, Lichtenbergstra{\ss}e 1, D-85747 Garching, Germany}
\affiliation{J\"{u}lich Centre for Neutron Science, Forschungszentrum J\"{u}lich, Outstation at Institut Laue-Langevin, BP 156, 38042 Grenoble Cedex 9, France}

\date{\today}

\begin{abstract}

Neutron scattering experiments were performed to investigate magnetic order and magnetic excitations in ternary iron chalcogenide K$_{2}$Fe$_{4}$Se$_{5}$. The formation of a superlattice structure below 580 K together with the decoupling between the Fe-vacancy order-disorder transition and the antiferromagnetic order transition appears to be a common feature in the \emph{A}$_{2}$Fe$_{4}$Se$_{5}$ family. The study of spin dynamics of K$_{2}$Fe$_{4}$Se$_{5}$ reveals two distinct energy gaps at the magnetic Brillouin zone center, which indicates the presence of magnetic anisotropy and the decrease of local symmetry due to electronic and orbital anisotropy. The low-energy spin wave excitations of K$_{2}$Fe$_{4}$Se$_{5}$ can be properly described by linear spin wave theory within a Heisenberg model. Compared to iron pnictides, K$_{2}$Fe$_{4}$Se$_{5}$ exhibits a more two-dimensional magnetism as characterized by large differences not only between out-of-plane and in-plane spin wave velocities, but also between out-of-plane and in-plane exchange interactions.

\end{abstract}

\maketitle

 New excitement in research on iron-based superconductors has arisen recently due to the discovery of the new superconducting compound K$_x$Fe$_{2-y}$Se$_2$ with \emph{T}$_C$ above 30 K \cite{Guo}. Superconductivity with similar critical temperature has been found soon after in other isostructural \emph{A}$_x$Fe$_{2-y}$Se$_2$ compounds with \emph{A} = Rb, Cs and Tl \cite{Wang1,Krzton-Maziopa,Fang1}. The extraordinary characteristic of electronic band structure in this system is the absence of hole pockets at the zone center. This poses a strong challenge to the well accepted paradigm for Fe-based superconductors concerning the pairing symmetry and the nature of magnetism based on the Fermi surface nesting scenario \cite{Wen}. Besides, the $\sqrt{5}\times\sqrt{5}$ type of Fe-vacancy order was observed and it was accompanied by the formation of antiferromagnetic blocks of Fe spins \cite{Bao, Ye, Wang2}. The optimal composition of \emph{A}$_{2}$Fe$_{4}$Se$_5$ is suggested according to the observed $\sqrt{5}\times\sqrt{5}$ Fe-vacancy-order pattern. In addition to the $\sqrt{5}\times\sqrt{5}$ superstructure phase, detailed experiments using transmission electron microscopy, x-ray, and neutron scattering revealed the existence of a $\sqrt{2}\times\sqrt{2}$ superstructure phase in superconducting samples  \cite{ Wang3,Wang2}. It is suggested that superconductivity might be present in the $\sqrt{2}\times\sqrt{2}$ phase rather than the $\sqrt{5}\times\sqrt{5}$ phase, which has a Mott insulator ground state. Further experimental work provided evidence of the phase separation on nanoscopic length scales in \emph{A}$_x$Fe$_{2-y}$Se$_2$ compounds \cite{Ricci, Yuan, Li2}. A thorough understanding of the physical properties of the superconducting phase as well as the relation between the superconducting and the antferromagnetic phases is still needed.

Although the $\sqrt{5}\times\sqrt{5}$ Fe-vacancy-ordered phase is a non-superconducting phase, it does exhibit close relationship with the superconducting phase. Actually, the $\sqrt{5}\times \sqrt{5}$ Fe-vacancy-ordered phase appears inevitably in all \emph{A}$_x$Fe$_{2-y}$Se$_2$ bulk superconducting compounds. Given the fact that dynamic magnetism in Fe-based superconductors may play an important role in mediating superconductivity \cite{Johnston}, it is necessary to understand both static and dynamic magnetism in the insulating $\sqrt{5}\times\sqrt{5}$ Fe-vacancy-ordered phase. The spin wave of block antiferromagnetic \emph{A}$_{2}$Fe$_{4}$Se$_5$ phase has been examined theoretically by many groups \cite{You, FLu,Fang2,Ke}. Furthermore, time-of-flight inelastic neutron scattering experiments have also been performed to investigate the spin wave dispersion of insulating Rb$_{0.89}$Fe$_{1.58}$Se$_{2}$, (Tl,Rb)$_2$Fe$_4$Se$_5$, and superconducting Cs$_{0.8}$Fe$_{1.9}$Se$_{2}$ compounds \cite{Wang4, Chi, Taylor}. These experiments reveal that the antiferromagnetic next-nearest-neighbor couplings exhibit comparable strengths in both iron pnictides and iron chalcogenides \cite{Zhao1,Lipscombe}.

In contrast to the neutron time-of-flight technique, triple-axis neutron spectrometry can provide more accurate dynamic information in a given region of energy and momentum space. In the present work, we studied the magnetism, especially the low-energy spin wave dispersion, of a block antiferromagnetic K$_{2}$Fe$_{4}$Se$_5$ compound with both diffraction and triple-axis inelastic neutron scattering techniques. We find that the energy spectrum at the Brillouin zone center can be modeled with two different spin anisotropy gap  parameters, which can be interpreted as indications for a breaking of local symmetry. A detailed analysis of the spin wave dispersion shows that the low-energy magnetic excitations of the antiferromagnetic block phase of K$_{2}$Fe$_{4}$Se$_5$ can be well described by a Heisenberg model with local magnetic exchange couplings extended to the third-nearest-neighbor. The larger energy band width observed in K$_{2}$Fe$_{4}$Se$_5$ is related to a stronger exchange coupling strength and higher antiferromagnetic transition temperature compared to other \emph{A}$_x$Fe$_{2-y}$Se$_2$ compounds.

The single crystals of K$_{2}$Fe$_{4}$Se$_{5}$ were grown by the Bridgman method as previously reported \cite{Landsgesell}. Single-crystal neutron diffraction and inelastic neutron scattering measurements were performed on the thermal-neutron two-axis diffractometer D23 and triple-axis spectrometer IN8 at the Institut Laue Langevin (Grenoble, France).  The crystal we used for the neutron scattering measurements was in the shape of a cylinder with a total mass of 3.9 g. For the diffraction measurement at D23, a Cu (200) monochromator was selected to produce a monochromatic neutron beam with wavelength of 1.28 $\buildrel_\circ \over {\mathrm{A}}$. For the inelastic neutron scattering measurements at IN8, pyrolitic graphite PG(002) was chosen as analyzer, while Si(111) or Cu(200) was selected as monochromator. The fixed final neutron energy of \emph{E$_f$} = 14.66 or 34.83 meV was used depending on different momentum and energy ranges. For convenience, in this paper we will always describe the neutron scattering data in the high symmetry tetragonal \emph{I4/mmm} space group notation with momentum transfer wave vectors \textbf{Q} = (\emph{q$_x$},\emph{q$_y$},\emph{q$_z$}) (in units of $\buildrel_\circ \over {\mathrm{A}}$$^{-1}$) at position (\emph{HKL}) = (\emph{q$_x$a}/2$\pi$, \emph{q$_y$b}/2$\pi$, \emph{q$_z$c}/2$\pi$) in reciprocal lattice units, where \emph{a} = \emph{b} = 3.898 $\buildrel_\circ \over {\mathrm{A}}$ and \emph{c} = 14.121 $\buildrel_\circ \over {\mathrm{A}}$ at \emph{T} = 2 K. The K$_{2}$Fe$_{4}$Se$_{5}$ single crystal was aligned in the scattering plane defined by the orthogonal vectors (2 1 0) and (0 0 1), in which spin wave excitations along main symmetry directions in the magnetic Brillouin zone can be surveyed.

\begin{figure}
\includegraphics[width=8.5cm,height=12.5cm]{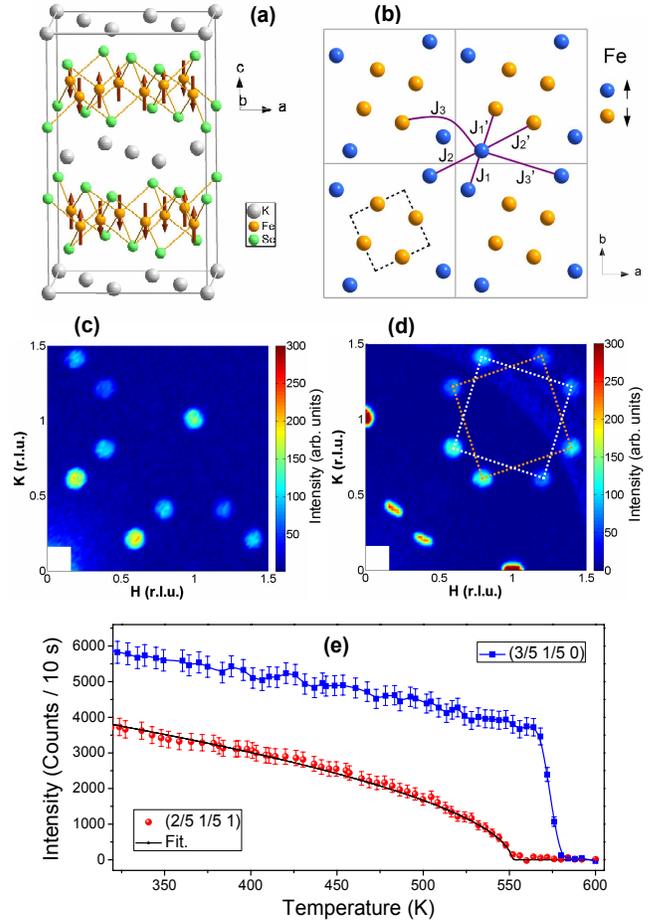}
\caption{\label{fig:epsart} (Color online) (a) Crystal and magnetic structure of K$_{2}$Fe$_{4}$Se$_{5}$  insulating phase. Gray lines highlight the low symmetry tetragonal unit cell. (b) Vertical view of antiferromagnetic spin block in Fe layer. The $\sqrt{5}\times\sqrt{5}$ superlattice unit cell is marked by gray bonds, while the dotted lines mark the high-temperature \emph{I4/mmm} unit cell. The blue and yellow spheres denote Fe atoms with spin-up and spin-down moments. The exchange couplings ($J_1$, $J_2$, $J_3$, $J_{1}^{\prime}$, $J_{2}^{\prime}$ and $J_{3}^{\prime}$) indicate intra- and inter-block exchange interactions. (c) and (d) Experimental contour map in the first quadrant of $(HK0)$ and $(HK1)$ plane in reciprocal space, where nuclear reflections and magnetic reflections from both chiral domains are observed. (e) Temperature dependence of integrated intensities of (3/5 1/5 0) nuclear and (2/5 1/5 1) magnetic superstructure reflections. }
\end{figure}

\begin{figure}
\includegraphics[width=8.5cm,height=7.5cm]{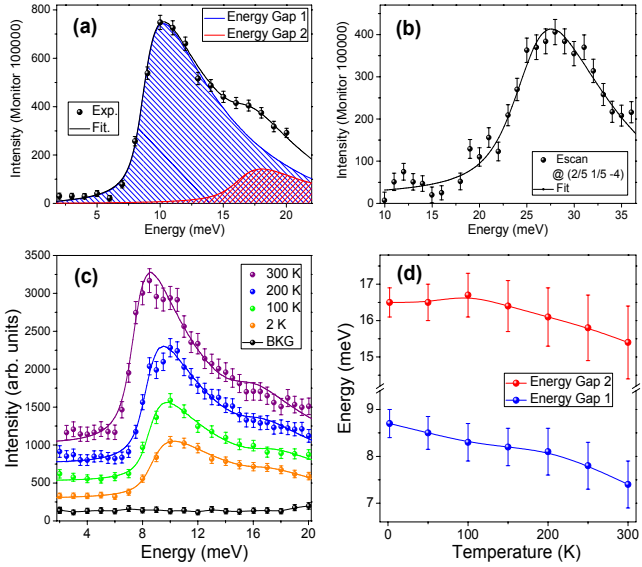}
\caption{\label{fig:epsart} (Color online) (a) Energy scan at the antiferromagnetic wave vector \textbf{Q} = (2/5 1/5 -3) at 2 K. The solid line is the fitting result with the dispersion model convoluted with the instrument resolution function. The spectral weights arising from two energy gap terms are highlighted as the shaded areas under the curve. (b) Energy scan at the zone boundary with wave vector (2/5 1/5 -4) at 2 K. (c) Temperature dependence of energy scan at wave vector \textbf{Q} = (2/5 1/5 -3). Solid lines indicate fitting results where corrected Bose population factor is also taken into account. Background scattering data (BKG) at an arbitrary wave vector is also shown for comparison. The curves are shifted upward to clearly show the change in energy gap features. (d) Variation of the two gap energies in the investigated temperature range. }
\end{figure}

The block antiferromagnetic structure [Fig. 1(a) and (b)] with the formation of the $\sqrt{5}\times\sqrt{5}$ superlattice has been suggested to describe the low-temperature structure for \emph{A}$_{2}$Fe$_{4}$Se$_{5}$ (\emph{A} = K, Rb, Cs and Tl) systems \cite{Bao,Ye}. Each block is composed of four Fe spins and these spins aligned ferromagnetically within the block, while the spins of neighboring blocks are aligned antiferromagnetically. Our neutron scattering results again confirmed the block antiferromagnetic structure as indicated by the contour map measured in (\emph{HK0}) and (\emph{HK1}) plane of the \emph{I4/mmm} unit cell [Fig. 1(c) and (d)]. The magnetic structure refinement based on collected integrated intensities gives ordered moments of 3.2(3) $\mu$$_B$ and 2.7(3) $\mu$$_B$ at 2 and 300 K, respectively. In Fig. 1(e), the rapid increase of the intensity of (3/5 1/5 0) structural peak below \emph{T}$_S$ = 580(3) K indicates the structural phase transition from the high-temperature Fe-vacancy-disordered phase with \emph{I4/mmm} symmetry into the low-temperature Fe-vacancy-ordered phase with \emph{I4/m} symmetry. The temperature variation of the integrated intensity of (2/5 1/5 1) magnetic peak can be fit with an empirical power law \emph{I} $\propto$ (1-\emph{T}/\emph{T}$_N$)$^{2\beta}$, which yields \emph{T}$_N$ = 553(3) K and a critical exponent $\beta$ of 0.27(1), as represented by the solid line in Fig.1 (e).

In Fig. 2(a), the energy scan at the Brillouin zone center \textbf{Q} = (2/5 1/5 -3) demonstrates a clear energy gap of magnetic excitation due to single ion anisotropy. The significant spectral signal is observed to extend up to 20 meV after substraction of background that was obtained by a comparable scan at \textbf{Q} = (0.7 0.35 -3). The existence of an energy gap is further demonstrated by \textbf{Q}-scans along two principal momentum directions below and above the gap energy as shown in Fig. S1(a) and Fig. S2(a). To determine the accurate gap energy, the observed energy scan data was used to fit within the framework of empirical spin wave relation convoluted with instrumental resolution function \cite{Tennant, McQueeney}. Interestingly, the data in the whole investigated energy range cannot be properly fitted with only one energy gap parameter. A second energy gap parameter has to be introduced. As a result, the analysis leads to two gap energies of 8.7(3) and 16.5(3) meV.  A similar feature was recently reported for BaFe$_2$As$_2$ where inelastic polarized neutron scattering studies lead to the observation of a strongly anisotropy spin excitation and the occurrence of two gap energies \cite{Qureshi}. The observed strong in-plane single ion anisotropy in BaFe$_2$As$_2$ demonstrated the important role of orbital degrees of freedom in the iron pnictides. Regarding \emph{A}$_{2}$Fe$_{4}$Se$_{5}$, it was found that the magnetic exchange energy will be minimized and the electron correlation will be enhanced by the presence of vacancy order in the $\sqrt{5}\times\sqrt{5}$ superlattice phase. Furthermore, a particular orbital order pattern will be present and it will result in the breaking of local fourfold symmetry on the Fe site \cite{Lv}. The observed two energy gap features in K$_{2}$Fe$_{4}$Se$_{5}$ might be an indication of the reduction of local symmetry due to electronic and orbital anisotropy. Despite the two energy gaps, only the one dominating the spectrum was used for the subsequent spin wave analysis. It is also noteworthy that the gap energy values obtained in K$_{2}$Fe$_{4}$Se$_{5}$ are of the same order of magnitude as the ones of various iron arsenides, in spite of quite large differences in ordering temperature and magnitude of Fe moments \cite{Park}.

Similar fitting procedures are performed to fit energy scan data observed at different temperatures [Fig. 2(c)]. The obtained two spin gap values as a function of temperature are shown in Fig. 2(d). Both energy gaps exhibit similar temperature dependence in spite of their different spectral weights. Narrowing of the energy gaps, which demonstrates the reduction of spin anisotropy, is clearly observed upon the increase of temperature from 2 to 300 K. Apart from the spin gap at the zone center, the Brillouin zone boundary is also examined. Fig. 2 (b) shows an energy scan at \textbf{Q} = (2/5 1/5 -4) located at the magnetic Brillouin zone boundary. The zone boundary energy is deduced to be 25.3(3) meV and is almost three times larger than the gap value at zone center. Spin wave energies at both zone center and zone boundary will help us to model the spin wave dispersion relationship in K$_{2}$Fe$_{4}$Se$_{5}$ as discussed in the following text.

\begin{figure}
\includegraphics[width=8.5cm,height=9cm]{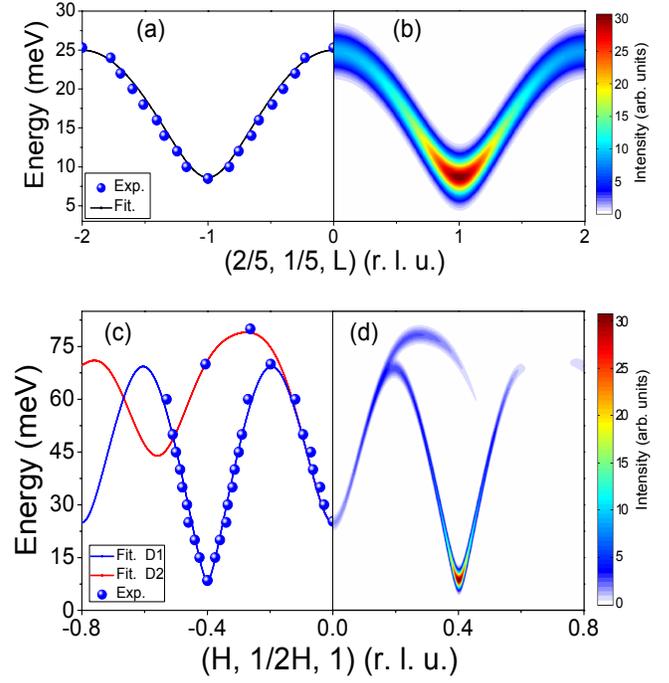}
\caption{\label{fig:epsart} (Color online) (a) Spin wave dispersion relation along \emph{L} direction as deduced from \textbf{Q}-scans at constant energy. The solid line is the fitted dispersion with a Heisenberg model as described in the text. (b) Variation of dynamic spin structure factors. (c) Spin wave dispersion along (2\emph{H} \emph{H} 0) direction. Solid lines are the fitted spin wave dispersions originating from two different structural domains (D1 and D2) with equal populations. (d) Variation of dynamic spin structure factors along (2\emph{H} \emph{H} 0) momentum space direction around \textbf{Q} = (2/5 1/5 1).}
\end{figure}

To determine the spin wave dispersion in the block antiferromagnetic K$_{2}$Fe$_{4}$Se$_{5}$ phase, constant energy measurements around the zone center \textbf{Q} = (2/5 1/5 l) were performed along two high symmetry directions in the Brillouin zone, i.e. (0 0 \emph{L}) and (2\emph{H} \emph{H} 0) directions (Fig. S1 and Fig. S2). The dispersion relations extracted from the measured scans are plotted as spheres in Fig. 3 (a) and Fig. 3 (c). Given the fact that K$_{2}$Fe$_{4}$Se$_{5}$ is an antiferromagnetic insulator with a rather large ordered Fe moment, we analyze spin wave dispersion data in linear spin wave approximation with Heisenberg Hamiltonian
\begin{equation}
H = \frac{1}{2}  \sum_{i,j}  J_{ij} S_i \cdot S_j -D_s \sum_{i} S^{2}_{i,z}
\end{equation}

\noindent where \emph{J}$_{ij}$ denote both in-plane coupling and out-of-plane coupling constants (\emph{J}$_c$), while \emph{D}$_s$ is the uniaxial single-ion anisotropy constant.

As illustrated in Fig. 1(b), eight Fe spins are distributed in the magnetic unit cell of \emph{I4/m} symmetry, thus one two-fold degenerate acoustic mode at lower energy together with three two-fold degenerate optical modes at higher energy are expected. As observed in isostructural Rb$_{0.89}$Fe$_{1.58}$Se$_{2}$ and (Tl,Rb)$_2$Fe$_4$Se$_5$ by time-of-flight neutron scattering measurements, all three optical modes are located at energies higher than 100 meV \cite{Wang4, Chi}. In our work, a gapped acoustic mode is measured accurately, but unfortunately the optical modes could not be reached because of the kinematic limit and low neutron flux at high transfer energy. The acoustic mode is found mainly due to the antiferromagnetic interaction between ferromagnetic blocked spins and it is mainly determined by $J_{1}^{\prime}$, $J_{2}^{\prime}$, $J_3$, $J_c$ and $D_s$ within the $J_{1}$$\text{-}$$J_{2}$$\text{-}$$J_{3}$ model. For instance, the energy gap at the zone center can be expressed as $\Delta_S$ = $S$$\sqrt{D_s(2J_{1}^{\prime}+4J_{2}^{\prime}+4J_{3}+D_{s})}$, while the spin wave energies at the zone boundary \textbf{Q} = (2/5 1/5 2) and at \textbf{Q} = (0 0 1) are found to be \emph{E}$_{ZB}$ = $S$$\sqrt{[2(J_{1}^{\prime}+2J_{2}^{\prime}+2J_{3}) +D_{s}] (4J_{c}+D_{s})}$. Therefore, we adopted the exchange parameters $SJ_1$ = -36(2) and $SJ_2$ = 12(2) meV from isostructural Rb$_{0.89}$Fe$_{1.58}$Se$_{2}$ \cite{Wang4} and fit our observed spin wave dispersion data with $J_{1}^{\prime}$, $J_{2}^{\prime}$, $J_3$, $J_c$ and  $D_s$ as variables. The spectra originated from two different chiral domains were also taken into account. The fitting results exhibit reasonable agreement with experimental data and yield exchange parameters as $SJ_{1}^{\prime}$ = 17(2) , $SJ_{2}^{\prime}$ = 19(2), $SJ_3$ = 12(2) , $SJ_c$ = 0.88(8) and  $SD_s$ = 0.46(6) meV.

\begin{figure}
\includegraphics[width=8.5cm,height=7cm]{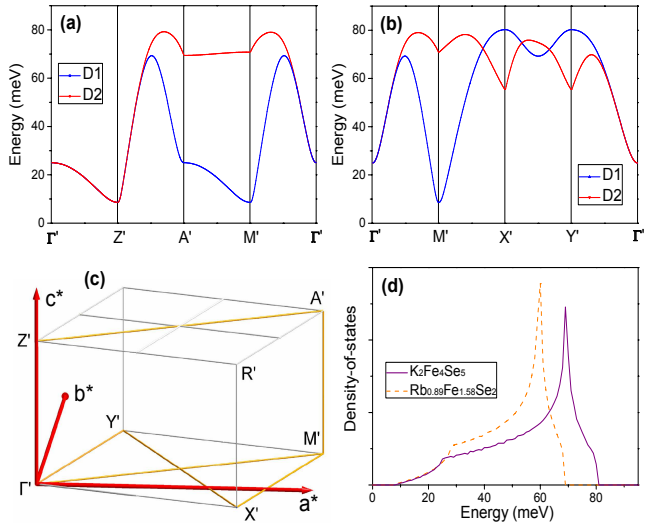}
\caption{\label{fig:epsart} (Color online) (a) and (b) Low energy spin wave branches along the selected high symmetry directions in the magnetic Brillouin zone. D1 and D2 represent spin waves originated from two different chiral domains. (c) Illustration of the defined high-symmetry points in the Brillouin zone. $\Gamma$$'$ = (0, 0, 2$\pi$/\emph{c}); M$'$ = (4$\pi$/5\emph{a}, 2$\pi$/5\emph{a}, 2$\pi$/\emph{c}); X$'$ = (3$\pi$/5\emph{a}, -$\pi$/5\emph{a}, 2$\pi$/\emph{c}); Y$'$ = ($\pi$/5\emph{a}, 3$\pi$/5\emph{a}, 2$\pi$/\emph{c}); Z$'$ = (0, 0, 4$\pi$/\emph{c}); A$'$ = (4$\pi$/5\emph{a}, 2$\pi$/5\emph{a}, 4$\pi$/\emph{c}). (d) Calculated low energy spin wave density-of-states for isostructural compounds K$_{2}$Fe$_{4}$Se$_{5}$ and Rb$_{0.89}$Fe$_{1.58}$Se$_{2}$. }
\end{figure}

In the $\sqrt{5}\times\sqrt{5}$ vacancy ordered \emph{A}$_2$Fe$_4$Se$_5$ phase, the nearest and the next-nearest neighbor interactions ($J_{1}$, $J_{1}^{\prime}$, $J_{2}$, $J_{2}^{\prime}$) are undoubtedly significant to describe its spin wave behaviors. However, the importance of the third-nearest-neighbor interaction ($J_{3}$) is still a matter of controversy \cite{Wang4, Chi}. In the present work, we found that the influence of $J_{3}$ reflected mainly in the change of the zone boundary energy of the acoustic branch. Based on the analysis on our own data as well as the results presented in Ref. \cite{Wang4, Chi}, we argue that the third-nearest-neighbor exchange interaction $J_{3}$ also plays a role in determination of spin dynamics in \emph{A}$_2$Fe$_4$Se$_5$. By considering the interactions extended to the third-nearest-neighbor, variations of the dynamic spin structure factor \emph{S}(\textbf{Q},$\omega$) are obtained and presented in Fig. 3(b) and (d) to illustrate the intensity distribution and the individual contribution from two different domains. It is clearly shown that the dynamic structure factors from two domains exhibit closeness and even overlap in certain \textbf{Q} range in the (2\emph{H} \emph{H} 0) direction.

The spin wave dispersion of the acoustic modes along all high symmetry directions is calculated by using the obtained exchange parameters, as shown in Fig. 4(a) and (b). The spin wave density-of-states (SWDOS) of K$_{2}$Fe$_{4}$Se$_{5}$ was also obtained by the summation over all wave vectors in the Brillouin zone, as plotted in Fig. 4(d). The density-of-states directly reflect the distribution of spin wave energies. Similar to pnictides, SWDOS of K$_{2}$Fe$_{4}$Se$_{5}$ exhibits sharp peaks or anomalies at energies of van Hove singularities \cite{Applegate, Johnston2}. The SWDOS of Rb$_{0.89}$Fe$_{1.58}$Se$_{2}$, calculated using the exchange parameters provided in Ref. \cite{Wang4}, is also given in Fig. 4(d) for comparison. It can be seen that the energy bandwidth of K$_{2}$Fe$_{4}$Se$_{5}$ is larger than that of Rb$_{0.89}$Fe$_{1.58}$Se$_{2}$, which reflects stronger exchange interactions and a higher antiferromagnetic transition temperature.

The investigations of the low energy spin wave dispersions of the parent phase of iron pnictides, e.g. in BaFe$_2$As$_2$, CaFe$_2$As$_2$ and SrFe$_2$As$_2$, have revealed values of \emph{v}$_{c}$/\emph{v}$_{a}$ = 0.2-0.5 for the ratio between out-of-plane and in-plane spin wave velocities (\emph{v}$_{c}$/\emph{v}$_{a}$)  \cite{Johnston, Lynn, McQueeney}. Therefore, it was suggested that the parent phases of iron pnictides are three dimensional antiferromagnets rather than quasi two-dimensional antiferromagnets as layered cuprates, in which finite but small interlayer coupling exists \cite{Johnston3}. The spin wave dispersion of K$_{2}$Fe$_{4}$Se$_{5}$ at lower energy yield an in-plane velocity of \emph{v}$_{a}$ = 380(20) meV$\cdot$$\buildrel_\circ \over {\mathrm{A}}$ and an out-of-plane velocity of \emph{v}$_{c}$ = 50(10) meV$\cdot$$\buildrel_\circ \over {\mathrm{A}}$. The ratio of \emph{v}$_{c}$/\emph{v}$_{a}$ = 0.13(2) observed in K$_{2}$Fe$_{4}$Se$_{5}$ indicates that it possesses more two-dimensional character in magnetism than iron pnictides. The more two-dimensional magnetism in K$_{2}$Fe$_{4}$Se$_{5}$ is also reflected by the ratio between out-of-plane and in-plane exchange interactions. For instance a $\frac{J_{out-of-plane}}{J_{in-plane}}$ ratio of less than 3\% is obtained if we compare $J_c$ with nearest neighbor exchange interaction $J_1$ or effective block in-plane exchange parameter $J_{eff}$ = $\frac{1}{4}$($J_{1}^{\prime}+2J_{2}^{\prime}+2J_{3}$) as suggested in Ref.\cite{Wang4}. It is believed that superconductivity is favored in more two-dimensional magnetic system since large spin fluctuations might suppress the long range order and mediate superconductivity.

In summary, we have investigated both static and low energy dynamic magnetism of a block antiferromagnetic K$_{2}$Fe$_{4}$Se$_{5}$ single crystal by using elastic and inelastic neutron scattering techniques. K$_{2}$Fe$_{4}$Se$_{5}$ is found to undergo a Fe-vacancy order-disorder transition and an antiferromagnetic transition at \emph{T}$_S$ = 580(3) and \emph{T}$_N$ = 553(3) K, respectively. Additionally, we obtained the acoustic mode of spin wave dispersions and it was properly fitted by using a linear spin wave model based on $J_{1}$$\text{-}$$J_{2}$$\text{-}$$J_{3}$ Heisenberg exchange couplings. Moreover, we found two well separated energy gaps at the magnetic Brillouin zone center. The breakdown of local tetragonal symmetry due to the emergence of orbital order might be responsible for the appearance of two energy gaps.

T.W. thanks the Deutsche Forschungsgemeinschaft for financial support under the DFG Priority Program 1458.

\appendix

\end{document}